\documentclass[journal]{IEEEtran}
\usepackage{graphicx}  
\usepackage{url}      
\usepackage{amsmath}                           
\usepackage{array}

\begin{document}
\title{Waveguide Model for Thick Complementary Split Ring Resonators}%
\author{\IEEEauthorblockN{L. M. Pulido-Mancera, J. D. Baena}\\ 
\IEEEauthorblockA{Group of Applied Physics, Physics Department, Universidad Nacional de Colombia
\\lmpulidom@unal.edu.co}}
\markboth{2014 IEEE International Symposium on Antennas and Propagation and USNC-URSI National Radio Science Meeting}{\MakeLowercase{\textit{et al.}}: Bare Demo of IEEEtran.cls for Journals}

\maketitle

\begin{abstract}
This paper presents a very simple analytical model for the design of Frequency Selective Surfaces based on Complementary Split Ring Resonators (CSRR) within the microwave range. Simple expressions are provided for the most important geometrical parameters of the model, yielding an accurate description of the CSRR resonance frequency and avoiding full-wave numerical simulations. Besides, a qualitative description of the band-pass filter behavior of these structures is described, considering its high quality factor Q.
\end{abstract}

\section{Introduction}

Frequency Selective Surfaces (FSSs) are filters for electromagnetic waves in free space, which often are considered plane waves. Last years, metamaterial based FSSs have received much attention because of the advantage of a small unit cell compared to the wavelength, which means that secondary grating lobes are avoided.\cite{Metamaterials-Book} It was demonstrated by Falcone and co-workers \cite{Babinet-Baena} that a periodical 2D array of Split Ring Resonator (SRR) behaves as a stop band filter and, according to the Babinet’s principle, its complementary screen (CSRR) behaves as a band pass filter. Both the bandstop and the passband are typically narrow, so that the quality factor (Q) is relatively big.

For the 2D case, the resonance frequency of the SRR and CSRR can be described by using an equivalent LC-circuit model \cite{LC-circuit-Baena}. However, numerical simulations have shown that when the thickness of the SRR increases, the resonance frequency is not the same as the complementary CSRR, the Babinet's principle is no longer satisfied. \cite{APS-2013}. LC-circuit models can still be applied for thick SRR resonators, but to thick CSRR resonators a different perspective is needed.

Transmission line models describe the response of thick periodic metallic structures, and reduce the problem to the computation of the scattering parameters of a height-step discontinuity in a parallel plate waveguide \cite{thick-screens}. The incoming electromagnetic plane wave, described as a transmission line, must match the admittance in the surface of the FSS to be propagated along rectangular slits, considered a transmission line too.  The S-parameters depends on the equivalent admittances of the incident wave, the surface and the slits and the admittance-matching occurs only at certain frequencies, such that the S-parameters presents different peaks and dips. 

In this paper, we propose a simple explanation to the first two resonances found in the $S_{11}$ and $S_{21}$ parameters, when the thickness of the CSRR increases. This brief explanation will simplify the calculation of the geometrical parameters to design FSSs, avoiding numerical simulations.

\section{Analytical model}
\begin{figure}
\includegraphics[width=0.9\columnwidth]{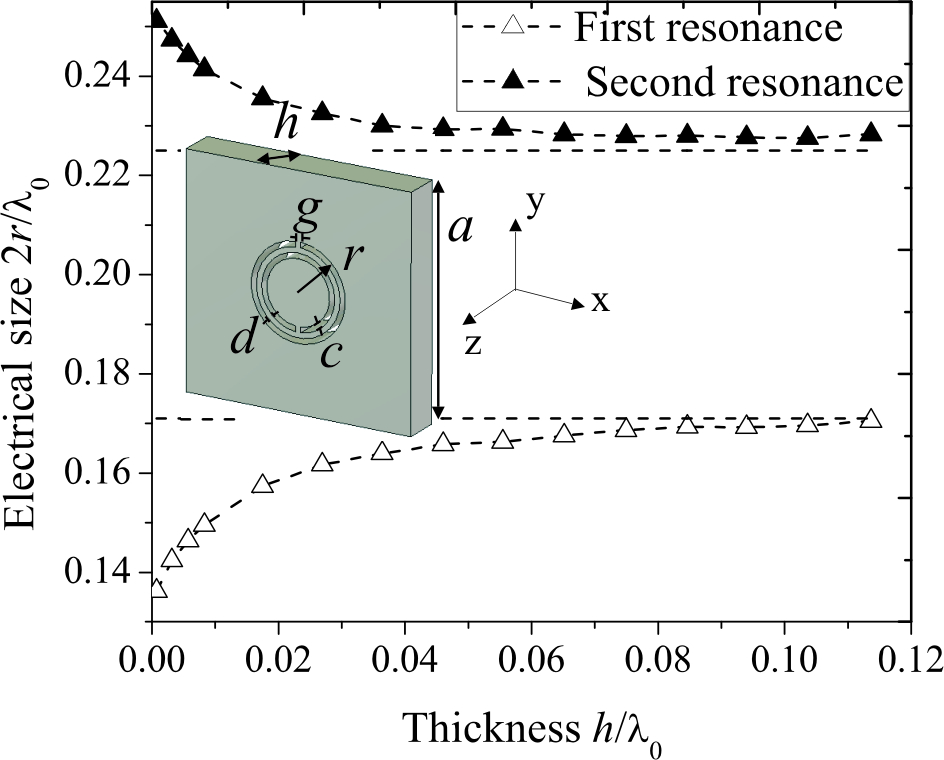} 
\caption{Electric size vs. Thickness $h$ normalized by resonance wavelength. The normalization shows that the CSRR converges to the cut-off frequency when the thickness increases. The geometrical parameters used in this simulation are $a=$43mm, $r=$9mm, $c=d=g=$1mm. $h$ vary within the range of 0-12mm. }
\label{normalization}
\end{figure}

Let us consider a periodic array of identical slits  with the CSRR geometry, made in a thick conducting plate. Figure \ref{normalization} represents a frontal view of the system under study, with $a$ being the period of the structure, $h$ the thickness of the metal plate, $c$ the slit width, $d$ the separation between slits, $g$ the gap and $r$ the external radius of the CSRR. An electromagnetic plane wave (TEM) impinges normally on the structure with its magnetic field oriented in the $x-$direction. 

In our proposed model, each slit of the CSRR can be seen as a rectangular waveguide bended with the shape of C. In this sense, the length $L$ of each waveguide is $2 \pi(r - c/2)- g$ and $2\pi (r - d - 3c/2)- g$ respectively. As a waveguide, the first frequency that allows the transmission of the wave is the cut-off frequency of the lowest mode (TE$_{10}$), when the resonance wavelength $\lambda_{0}$ is twice the length of the waveguide.

In order to validate this condition, we performed several simulations using \textit{CST-Microwave Studio Software} and we obtained the two primary resonances of the $|S_{11}|$ and $|S_{21}|$ parameters. As shown in Fig. \ref{normalization}, the $x$ axis is the thickness $h$ divided by $\lambda_{0}$, and the $y$ axis is the resonance frequency $f_{0}= c/\lambda_{0}$, rewritten in terms of the electrical size $2r/\lambda_{0}$. As the thickness increases, each curve tends to a fixed value that corresponds to the ratio between the diameter and the length of each waveguide, such that $2 r/ (2 \pi r -c/2)-g =$0.171 and $2 r/ 2 \pi (r - d - 3c/2)-g =0.225$.

Figure \ref{normalization} is very useful for the FSS design because it gives us information in order to estimate accurately $\lambda_{0}$: it is necessary to choose a thickness greater than $h= 0.006\lambda_{0}$ and $L= \lambda_{0}/2$, with $L$ being dependent on $r$, $g$, $d$ and $c$. Other simulations that are not presented in this paper show that the variance of $c$ and $g$ is small, indeed the cut-off frequency in a waveguide does not depend on its height. Thus, the problem of filter design is reduced to estimate the value of $r$. In addition, given that the first and second resonances operate independently, it is possible to design a CSRR adjustable to two different frequencies using a single unit cell geometry and just varying $d$. In this sense, thick CSRR structures not only has the advantage of small electrical size but also adjustable response to two frequencies working separately.

\section{S-parameters for CSRR}

\begin{figure}
\includegraphics[scale=1]{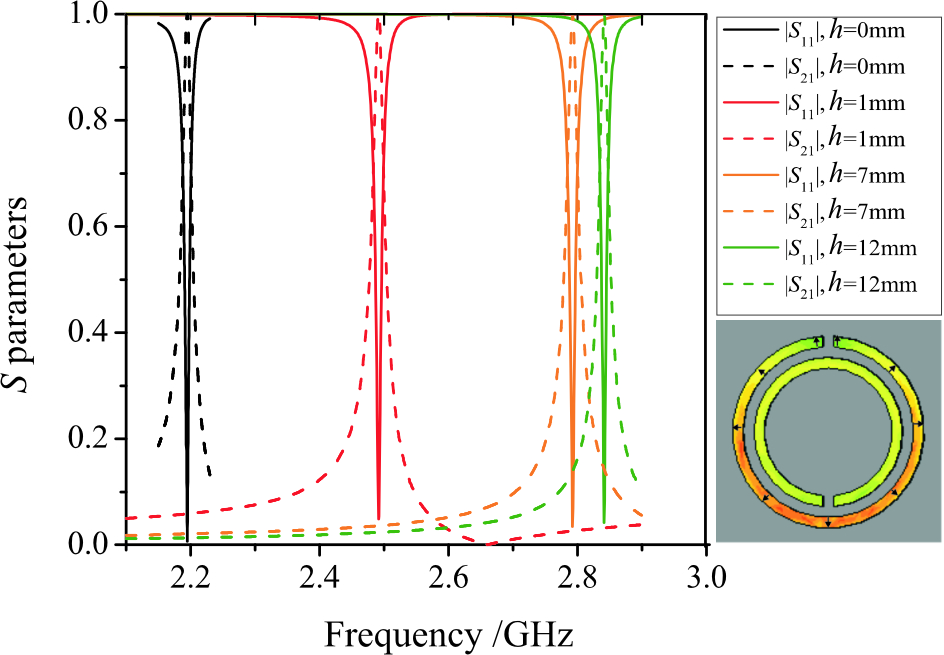} 
\includegraphics[scale=1]{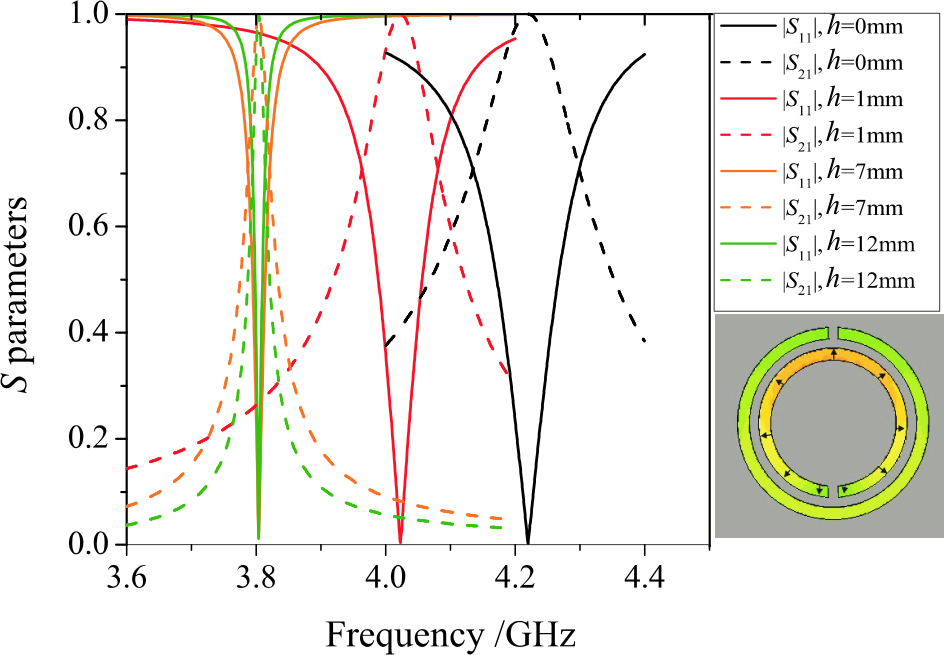} 
\caption{S-parameters for different values of thickness using full-wave simulation in CST-Studio. a.) first resonance, b.) second resonance. Color curves varies for each value of thickness. Reflection S11 (solid line)is close to 0 and transmission S21 (dashed line) is close to 1 at resonance.}
\label{transmission-peaks}
\end{figure}

The transmission curves show how the first resonance frequency $f_{1}$ tends to a fixed value as the thickness $h$ increases. Figure \ref{transmission-peaks}a. shows the S parameters corresponding to the first resonance. As shown, $f_{1}$ increases but its bandwidth remains constant, thus; the quality factor $Q= f/\Delta f$ increases, as it had been previously demonstrated in \cite{APS-2013}. Likewise, Q increases in the second resonance $f_{2}$, but in this case the increment is related with the bandwidth becomes narrower in spite of the fact that $f_{2}$ decreases. 

By analyzing the electric field distribution and the surface current induced on the metallic surface only at the resonance frequency, it is possible to observe how each slot of the CSRR resonates independently when the structure is thick. Thus, within the frequency range between 2.2-2.9 GHz, only the outer slit allows the wave transmission while the internal slot acts as a metallic screen, Fig. \ref{transmission-peaks}a. Likewise, within the frequency range between 3.8 - 4.4 GHz, the internal slot fully transmits while the outer reflects almost completely, Fig.\ref{transmission-peaks}b. When the surface is thin, the coupling between the two slots begins to be important so the resonance cannot be predicted with our proposed model.

Another interesting factor is the electric field direction inside each slit because it satisfies the boundary conditions of a bent rectangular waveguide. In fact, this distribution of the electric field was the key factor that allowed us to describe the geometry of CSRR rectangular widths in an analogous way to rectangular waveguides that transmit at the cut-off of the fundamental mode.

\section{Conclusions}
This model is very useful to find the geometric parameters necessary for to design a structure made of CSRR geometry that completely transmits at a fixed frequency. The most important parameters are the ratio, $r= c/4\pi f_{res}$, and the thickness $h=0.06\lambda_{res}$. Besides, for thick structures the resonances work independently for each slit of the CSRR, which allow us to design tunable filters with high Q for two different frequencies within the same unit cell.

\end{document}